\RequirePackage{fix-cm}
\documentclass[smallextended]{svjour3}       
\smartqed  
\usepackage{graphicx}
\usepackage{listings}
\usepackage{xcolor}
\usepackage{amssymb}
\usepackage{amsmath}
\usepackage{rotating}
\usepackage{hyperref}
\usepackage{multirow}
\usepackage{array} 
\usepackage{tabularx}
\usepackage[authoryear]{natbib}

\newcommand{\code}[1]{\texttt{#1}}

\def\revise#1#2{{\color{red}{\small #1}}{\color{red}{\mbox{$\Rightarrow$}}}{\color{blue}{#2}}}
\newcommand{\reviseno}[2]{{#2}}
\renewcommand{\revise}[2]{#2}

\lstdefinestyle{java}{ 
    language=java,
    basicstyle=\footnotesize\ttfamily, 
    breakatwhitespace=false, 
    breaklines=true, 
    captionpos=b, 
    commentstyle=\color[rgb]{0,0.6,0}, 
    deletekeywords={}, 
    firstnumber=1, 
    frame=none, 
    frameround=tttt, 
    keywordstyle={[1]\color{blue!90!black}},
    keywordstyle={[3]\color{red!80!orange}},
    morekeywords={}, 
    numbers=left, 
    numbersep=3pt, 
    numberstyle=\tiny\color[rgb]{0.5,0.5,0.5}, 
    rulecolor=\color{black}, 
    showstringspaces=false, 
    showtabs=false, 
    stepnumber=1, 
    tabsize=2, 
    backgroundcolor=\color{white}
}

\begin{document}

\title{Can defects be fixed with weak test suites?}
\subtitle{An analysis of 50 defects from Defects4J}


\author{Jiajun Jiang \and Yingfei Xiong}

\institute{Jiajun Jiang \at
              Key Laboratory of High Confidence Software Technologies (Peking University), MoE 
              \\Institute of Software, EECS, Peking University, Beijing, 100871, China\\
              \email{jiajun.jiang@pku.edu.cn}           
           \and
           Yingfei Xiong \at
              Key Laboratory of High Confidence Software Technologies (Peking University), MoE 
              \\Institute of Software, EECS, Peking University, Beijing, 100871, China\\
              Tel.: +86-10-62757008\\
              Fax: +86-10-62751792\\
              \email{xiongyf@pku.edu.cn}
}

\date{Received: date / Accepted: date}

\maketitle

\begin{abstract}
  Automated program repair techniques, which target to generating
  correct patches for real world defects automatically, have gained a
  lot of attention in the last decade. Many different techniques and
  tools have been proposed and developed. However, even the most
  sophisticated program repair techniques can only repair a small
  portion of defects while producing a lot of incorrect
  patches\revise{, giving a low recall and a low precision}{}.
  \revise{ To better understand the upper bound of automated program
    repair, we conducted a human involved study where a human
    developer fixes the bugs under the setting of automatic program
    repair tools. In our study, the human developer fixed 84\% of the
    defects and achieved a precision of 89.4\%, which suggests that
    there exist a lot of spaces for automatic repair techniques to
    improve. After analyzing the fixing process, we further summarized
    seven fault localization strategies and seven patch generation
    strategies utilized by our human developer. Finally, we discuss
    possible directions to improve current automated program repair
    techniques based on the findings. }{ 
A possible reason for this
    low performance is that the test suites of real world programs are
    usually too weak to guarantee the behavior of the program. To understand to what extent defects can be fixed
    with weak test suites, we analyzed 50 real world defects
    from Defects4J, in which we found that up to 84\% of them could be
    correctly fixed. This
    result suggests that there is plenty of space for current
    automated program repair techniques to improve. Furthermore, we
    summarized seven fault localization strategies and seven patch
    generation strategies that were useful in localizing and fixing
    these defects, and compared those strategies with current
    repair techniques. The results indicate potential directions to improve
    automatic program repair
    in the future research. 
%
%
 }
    
\keywords{Bug fixes  \and Manual repair \and Software maintenance}
\end{abstract}

\section{Introduction}
\label{sec:intro}
Automated program repair techniques, which automatically generate
patches for defects in programs, have gained a lot of attention in the
last decade. A typical automatic program repair technique takes a
program and a set of tests as input, where at least one test is failed
by the program, and generates a patch that will fix the defect.
Different automated techniques and tools have been proposed. These
tools generate a patch through techniques such as directed random
search~\citep{le2012genprog,long2015staged},
templates~\citep{kim2013automatic}, component-based program
synthesis~\citep{nguyen2013semfix,mechtaev2015directfix,mechtaev2016angelix},
program transformation from
examples~\citep{gao2015fixing,long2016genesis,icse17-transformation} and machine
learning~\citep{long2016automatic}, incorporate fault-localization
approaches such as spectrum-based fault
localization~\citep{abreu2007accuracy,abreu2009spectrum}, predicate
switching~\citep{zhang2006locating}, and angelic
debugging~\citep{chandra2011angelic}, and utilize information such as
testing results~\citep{le2012genprog,marcote2016nopol}, existing
patches~\citep{kim2013automatic,long2016automatic,gao2015fixing},
invariants~\citep{DBLP:conf/sosp/PerkinsKLABCPSSSWZER09}, existing
source code~\citep{xiong2016precise},
bug report text~\citep{liu2013r2fix}, and comments~\citep{xiong2016precise}. 


\revise{
Despite these efforts, in practice even the most sophisticated program
repair techniques can only repair a small portion of defects while
producing a large number of incorrect patches, giving a low recall and
precision. In this paper we define precision as the portion of the
defects that are fixed by the first generated patch among all defects
that a patch is generated, and define recall as the portion of defects
that are repaired by the first generated patch among all defects. For
example, Prophet~\citep{long2016automatic} and
Angelix~\citep{mechtaev2016angelix}, two newest approaches on the C
language, have precisions of 38.5\% and 35.7\% and recalls of 14.3\%
and 12.2\% on the GenProg benchmark~\citep{EightDollar}, respectively.
The newest approach on Java, ACS~\citep{xiong2016precise}, has a
precision of 78.3\% and a recall of 8.0\% on the Defects4J
benchmark~\citep{just2014defects4j}.
}{
Despite these efforts, in practice even the most sophisticated program
repair techniques can only repair a small portion of defects while
producing a large number of incorrect patches. For
example, Prophet~\citep{long2016automatic} and
Angelix~\citep{mechtaev2016angelix}, two newest approaches on the C
language, can only fix 14.3\% and 12.2\% of the defects on GenProg
benchmark~\citep{EightDollar}, while producing incorrect patches for
other 22.8\% and 22.0\% defects, respectively. 
The newest approach on Java, ACS~\citep{xiong2016precise}, can only
fix 8.0\% defects on Defects4J benchmark~\citep{just2014defects4j}
while producing incorrect patches to other
2.2\% defects.}

\revise{}{An often attributed reason for this low performance,
  especially the large number of incorrect patches, is that the test
  suites of real world programs are usually weak. the incomplete
  specifications of real world programs. As mentioned above, automatic
  program repair techniques usually rely on tests to distinguish
  correct and incorrect patches. 
  However, as studied by \cite{qi15} and \cite{long2016automatic}, test
  suites in real world programs are often weak, and in a space of
  patches that pass all tests, there are often many more incorrect
  patches than correct patches. As a result, it is very difficult for
  automatic program repair techniques to distinguish incorrect patches
 and correct patches. Besides leading to more incorrect patches, weak
 test suites may also make it more difficult to locate correct
  patches. When we have a strong test suite or a formal specification,
  we can focus on the difference between the program and the
  specification to guide the search. However, when we have a weak test
  suite, the difference may be much larger and the guidance may be
  much more inefficient. 
  }

\revise{Since the current results are not very positive, it naturally raises a
    question: how much possible program defects can be repaired
    automatically? Or, more specifically, how much possible program
    defects can be repaired automatically under the current settings? This
    question is important because, if the performances of the current
    techniques are already close to the upper bound of defects that can be
    automatically repaired, we need to change the current problem
    setting, for example, by involving more information sources into the
    repair process, or by involving the developers interactively. On the
    other hand, if the performances of the current techniques are still
    far from the upper bound, we can focus on improving the current
    techniques.
}{
    Since the performance of current repair techniques are still
    limited, it naturally raises a
    question: is it possible to repair a large portion of defects with
    only weak test suites?
    This question is important because, if most of the defects cannot
    be fixed, we may change the problem settings of automatic program
    repair, e.g., asking the user to provide formal specifications
    of the programs. On the contrary, if most defects can be fixed, we
    can focus on improving the current techniques. 
  }

\revise{
To answer this question, we conducted a case study on a manual repair
process of real world defects. We randomly selected 50 defects from
Defects4J, a widely-used benchmark of real world defects in Java
programs. Then a human developer, who is the first author of the paper
and Ph.D. student with four-year's
experience in Java programming, attempted to manually repair these
defects under the same setting of automatic program repair techniques.
That is, the author is only presented with the program source code and
a set of tests, without any other information about the defects. The
human developer has no knowledge of the projects being fixed, either.
We documented the whole manual repair process and analyzed the
records.}{
To answer this question, we analyzed 50 defects randomly selected
from Defects4J, a widely-used benchmark of real world defects in Java
programs, to see how much possible these defects can be fixed.
In our analysis, a defect was considered as repairable if
and only if (1)
we could identify a possible root cause of the defect, (2) we could generate a patch
that tackles the root cause and passes all the tests, and (3) the patch is equivalent to the developer patch
provided by Defects4J.
}

\revise{ This case study could help us to understand the potential
  boundary of automatic defect repair. If we found many more defects
  can be fixed than current state-of-the-art approaches, it indicates
  that the lack of complete specification may not be the key limiting
  factor and we should focus on improving automatic repair techniques.
  Also, the study provides insights on how to improve automatic defect
  repair techniques. By comparing the current techniques with the
  manual defect repair process, we may find the weak point of current
  techniques and identify directions for improving the current
  techniques. }{ This study could help us to understand the potential
  of automatic defect repair and to improve current techniques. If a
  defect is considered as repairable in our analysis, there exists at
  least a manual process to obtain the patch for the defect. By
  decomposing and automating the manual process, we can potentially
  obtain an automatic method to repair the defect.
  Furthermore, if we found many more defects
  can be fixed than current state-of-the-art approaches, it indicates
  that weak test suites may not be the key limiting
  factor and we should focus on improving automatic repair techniques. 
}

During the analysis of \revise{the records}{those defects}, we focus on the following three
research questions:
\begin{itemize}
\revise{
    \item RQ1: What is the precision and recall of the manual repair process?}{
\item RQ1: How many of those defects can be fixed with weak test suites?}
\revise{
\item RQ2: How did the developer repair the defects?}{
\item RQ2: How those defects can be fixed?
}
\revise{
\item RQ3: How much can the manual repair process be automated?}{
\item RQ3: What is the implications of the analysis for improving
  automatic program repair techniques?
}
\end{itemize}



And our study has the following main results.
\begin{itemize}
\revise{
\item RQ1: The manual repair process achieved a precision of 89.4\% and a 
recall of 84.0\%, both of which significantly outperform the 
state-of-the-art repair techniques. This number indicates that 
there is still a lot of room for the automatic repair techniques to improve.
}{
\item RQ1: We found that at least 42 (84.0\%) of the 50 defects could
  be fixed with weak test suites. For the rest of the defects, 
  incomplete patches may be generated for 5
  (10.0\%) defects when relying on tests as specification, and the
  remaining 3 defects require domain-knowledge that is difficult to
  obtain from the program and the tests. The results indicate that
  current techniques have a lot of rooms for improvement and weak test
  suites may not be the key limiting factor
  for current techniques.
}
\revise{
\item RQ2: We identified seven strategies for fault localization and seven
  strategies for defect repair from the manual repair process.
  }{
\item RQ2: We summarized seven strategies 
for fault localization and seven strategies
for patch generation, which played an important role for
fixing these defects.
}
\item \revise{RQ3: We found that most strategies can be automated. Some of them
  have been utilized by existing approaches, but further improvements
  seem to be possible. The rest may lead to new repair techniques.
  This result shows the possibility of improving the \reviseno{recall}{performance} of
  existing techniques.}{
  RQ3: We found that many strategies have already been explored by
  current automatic program repair approaches. However, to fix those
  defects, current techniques still need to be improved. We compared
  the current techniques with each strategy and identified the
  concrete points that current techniques need to be improved. }
\end{itemize}

\revise{}{Please note that our results should not be interpreted as an
  upper bound on the performance of the automatic program repair
  techniques. If a defect was not repaired in our analysis, it may
  still be repairable by automatic repair approaches. For
  example, there may exist a method to fix the defect but we did not
  see it, or an automatic technique may rely on the computation power
  of the machine to obtain a patch while we as humans cannot. 
    }

The rest of the paper is organized as follows.
Section~\ref{sec:related} introduces the background of automatic program repair techniques and related empirical studies on defect repair. 
Section~\ref{sec:experiment} introduces the dataset and environment of our study and
Section~\ref{sec:result} analyzes the experiment result in detail to answer the research questions.
Section~\ref{sec:discussion} discusses the generalizability of our result and
Section~\ref{sec:conclusion} concludes this paper and introduces our future work.

\section{Background and Related Work}
\label{sec:related}

\subsection{Automatic Program Defect Repair}

As mentioned in the introduction, in a typical defect repair setting
the repair technique takes as input a program and a set of tests, where the
program fails at least one test, 
and produces as output a patch \revise{, which}{that}
is expected to repair the defect when applied to the program.
\revise{}{Since tests are used as primary tool to guarantee the correctness of
the patches, we call this setting as \emph{test-based program repair}.}

A key issue to evaluate the performance of repair tools is how to
determine the correctness of the generated patches. In the early
studies~\citep{le2012genprog,kim2013automatic} of automatic repair, a
patch is usually considered correct if the patched program passes all
the tests. In recent
studies~\citep{gao2015fixing,long2015staged,long2016automatic,mechtaev2016angelix,xiong2016precise},
a patch is usually considered as correct if it is semantically identical to
the patch produced by human. Note that neither approach can produce an
ideal measurement of correctness: the former may overstate the number
of correct patches (because the test suites may be too weak to
guarantee correctness) while the latter may understate the number of
correct patches (because a defect may be repaired in different ways).
However, as studied by \cite{qi15}, the former approach is
very imprecise for real world programs because the test suites are
usually weak. 
\revise{}{
Similarly, \cite{smith2015overfitting} studied that inadequate test suite
would lead to over fitting patches and suggested that repair techniques 
must go beyond testing to characterize functional correctness of patches.
}
As a result, in this paper we take the latter approach,
determining the correctness by the equivalence with human patches.

Many defect repair approaches follow a ``generate-and-validate''
approach. That is, these approaches first try to locate a likely patch
in a large patch space, and then validate the patch using all the
tests. 
There are two main challenges in the repair process. 
\revise{
    The first
is precision. As mentioned above, 
}{The first is to ensure the correctness of the generated patches. As
  mentioned above,}
the tests in the
real world programs are often not enough to guarantee correctness,
and thus it is difficult to ensure the correctness of the generated
patches. 
\revise{
The second is recall. Since
}{The second is to generate correct patches to a large number of
  defects. Since}
 the patches need to be validated
against all tests, the number of generated patches cannot be large. In
order to locate a small number of likely patches from the patch space,
current approaches cannot support a large patch space. 
As studied by \cite{long2016analysis} 
and \cite{zhong2015empirical}, most defects cannot be fixed by the
patch space considered in current approaches.

There are also defect repair approaches that use a different problem
setting. For example, some approaches assume that there exists a full
specification of the program~\citep{dqlose,wei2010automated}, and some
approaches consider a concrete class of defects such as memory
leaks~\citep{gao2015fixing} and deadlocks~\citep{Cai16}. These different problem settings are
not the focus of our paper.

\revise{
Besides precision and recall, there are other quality attributes, such as
maintainability~\citep{fry2012human} and
acceptability~\citep{tao2014automatically}. However, since precision
and recall are two basic quality attributes, in this paper we will
focus on precision and recall.
}{}

\subsection{Empirical Studies on Defect Repair}
%

There exist several empirical studies on defect repair. 
\cite{zhong2015empirical} studied the real bug fixes through
analyzing the commits of five open source projects. 
In their study, they analyzed the distributions of fault locations and
modified files. To investigate the complexity of fixing bugs, they
analyzed the data dependence among faulty lines. More concretely, they
analyzed the operations of bug fixes and how many of them related to APIs. 
Similarly, \cite{martinez2015mining}
studied the distribution of real bug fixes by analyzing a large number
of bug fix transactions of software repositories. In order to better
understand the natural of bug fixes, they classified those bug fixes
with different classification models. 
Besides, \cite{soto2016deeper} analyzed a great deal of bug-fixing
commits in Java projects aiming to provide a guidance for future
automatic repair approaches. 
\revise{In contrast to our study, their studies
focus on a large set of patches fixed under different settings, and
it is difficult to derive conclusions on the upper bound under the
current setting of automatic program repair.
Besides, their studies
focus on the result of bug fixes but not the process of bug fixes, and
thus it is difficult to derive knowledge on how the fixes are deduced. }
{In contrast to our study, their studies focus on the distribution of characteristics
    about defects and patches but not how these defects were fixed,
    thus it is difficult to derive conclusions on the repairability of
    the defects. 
}

Existing automatic program repair approaches have extracted templates
about defect repairs. \cite{kim2013automatic} have
proposed an automatic fixing approach, PAR. It generates patches for
defects by applying a set of templates predefined by human
developers. \cite{tan2016anti} predefined a set of
\textit{anti-patterns} to filter the undesirable patches generated by
other approaches. Compared with the strategies derived
from our analysis, the templates used in these approaches are mainly
syntactic templates derived from the changes, 
\revise{
while our strategies 
connect more on the process the human developer deduces the
patch. Furthermore, our result provides insights on how the human
developer gain confidence on the patch correctness. 
}{
while our strategies try to reason why the program failed from a developer
point of view and connect more on the process of how the patches
can be deduced.
}

There exist other human involved studies. \cite{tao2014automatically} has 
conducted a study that repairing real defects manually under the help of automatic program repair techniques. 
It is different with ours because they focus on how the generated patches help the developers rather than 
\revise{
    how the developers produce the patches.}{
    how patches can be derived.}
\revise{}{Several researchers studied the debugging process of human
  developers. 
\cite{joseph2013foraging} studied how human developers navigate
through the debugging
process and created a model for predicting the navigation process.
\cite{thomas2010hard} studied the questions developers ask during the
debugging process. \cite{emerson2015navigate} studied the factors
developers consider during the debugging process. 
Different from these studies, our study focus on analyzing the
repairability of defects rather than understanding how human developers
behave. 
}

\section{Dataset and Environment}
\label{sec:experiment}
We conducted our case study on the dataset
Defects4J~\citep{just2014defects4j}, which is a commonly-used
benchmark for automated program repair research. It 
consists of 357 defects from five projects and Table~\ref{tab:data}
lists the detail for each project. \revise{}{The test suites in Defects4J are
also known to be weak. There are usually only a few tests cover the
faulty place, and in many cases only the failed test cover the faulty
place. Existing studies~\citep{durieux2015automatic} also show that
program repair techniques such as jGenProg generate a lot of incorrect
patches on the benchmark.}

\revise{In our case study, }{Since the whole Defects4J is too large for manual analysis, }we randomly
selected ten defects from each project, and thus have a dataset of 50 defects.

\begin{table}[h]
    \begin{tabular}{llp{4.5cm}c}
        \hline\noalign{\smallskip}
        \textbf{ID} & \textbf{Project name} & \textbf{Description} &
            \textbf{\#Defects} \\
        \noalign{\smallskip}\hline\noalign{\smallskip}
        Chart   & JFreechart          & An open source framework for Java to create chart        & 26  \\
        Closure & Closure compiler    & A tool to optimize JavaScript source code                 & 133 \\
        Lang    & Apache commons-lang & A complement library for \code{java.lang} & 65\\
        Math    & Apache commons-math & A lightweight mathematics and statistics library for Java & 106\\
        Time    & Joda-Time           & A standard date and time library for Java & 27\\
        \noalign{\smallskip}\hline
    \end{tabular}
    \caption{Statistical for Defects4J Benchmark}
    \label{tab:data}       
\end{table}

\revise{In our case study, a human developer, who is the first author of this
paper and Ph.D. student with four-year's experience in Java
programming, attempted to repair the 50 selected defects. The human
developer had access to the source code of the projects, including the
passed and failed tests. The human developer had no prior knowledge of
the projects he was going to fix, nor did he have access to other
documentations of the projects beyond the source code. 
During the repair process, the human developer was allowed to access the
Internet, but was not allowed to obtain information specifically
related to a project under repair. In this way, the human developer
was put under the same setting as automated program repair tools.
}{
To understand how many defects can be repaired, we analyzed each defect
in the dataset to determine whether we can locate a correct patch for
the defect. Our analysis is performed under the following three
environment settings.
\begin{itemize}
\item We do not have prior knowledge of the Defects4J program. In
  other words, we do not know the expected behavior of the programs.
\item We only rely on the source code of the program to generate the
  patch, including both implementation code and testing code.
\item We do not use other resources such as program documents.
\end{itemize}
In this way, we put ourselves into the same environment setting as
most test-based program repair techniques. If we obtain the correct
patch for a defect under this setting, it indicates potential to fix
the defect automatically by decomposing and automating the manual
repair process.
}

\revise{}{More concrete, our analysis would classify the defects into
  \emph{repairable} and \emph{difficult to repair}, and the
  classification is based on the following steps
  for each defect. The first author of the paper, who is a Ph.D.
  student with four-year's experience in Java
programming, performed the analysis.
  \begin{itemize}
  \item Based on the implementation code and the testing code, we try
    to locate a possible root cause of the defect.
  \item We generate a candidate patch for the defect, and run all
    tests to validate the patch. If the patch does not pass all the
    tests, we restart from the first step.
  \item If the patch passes all tests, we further compare it with the
    developer's patch. If the two patches are equivalent,
    we regard the patch as correct and the defect as \emph{repairable},
    otherwise we regard the patch as incorrect and the defect as
    \emph{difficult to repair}.
  \item If we cannot obtain a patch that passes all tests within 5
    hours, we stop and consider the defect \emph{difficult to repair}.  
  \end{itemize}
}

\revise{}{
Here we decide the equivalence of two patches by semantic equivalence:
We say two patches are
equivalent only when we can transform one patch to the other one
by applying a series of semantics-preserving transformations.
\\
The detailed analysis of each defect is available online at \url{https://github.com/xgdsmileboy/Bug-Fixing-Records}.
}
\revise{
The repair process was
conducted on a 64bit OSX system with a 2.7 GHz Intel Core i5 processor
and 8G memory. The version of Java Runtime Environment is JavaSE 1.7
update 80. Besides, the human developer chose to use Eclipse Mars.1
Release (4.5.1), which he was familiar with, as the debugging tool. We
measured the time that the developer used for repairing each defect,
and the human developer was also asked to write down the whole repair
process after fixing each defect. Then we analyzed the records and
tried to answer the research questions.}{}

\newcounter{finding}
\newcommand{\finding}[1]{\refstepcounter{finding}
  \vspace{3.0pt}
  \fbox{
    \parbox{0.85\textwidth}{
      \textbf{Finding \arabic{finding}. } \emph{#1}
    }
  }
  \vspace{3.0pt}
}

\section{Results}
\label{sec:result}
In this section, we \revise{analyze the experiment result in
  detail}{present the result of our analysis} and answer
the research questions presented in Section~\ref{sec:intro}.


\subsection{RQ1: Defect-Analyzing Result}
\label{subsec:rq1}
\revise{Totally, the developer generated 47 patches for 47 defects, one for
    each defect. Then we validated
    the patches according to the standard patches provided by
    Defects4J~\citep{just2014defects4j}. The result is 42 out of them are
    completely correct, giving a recall of 84.0\% and a precision of 89.4\%.
}{
Among the 50 defects we analyzed, we classify 42 (84.0\%) defects as \emph{repairable}
and 8 defects as \emph{difficult to repair}. Table~\ref{tab:comparing}
shows the detailed data per each project as well as the comparison
with a set of existing program repair approaches. As we can see from the table, the
performance of existing program repair approaches can only repair a
very small portion of repairable defects, indicating a large room for
improvement. 
}

\revise{}{
\begin{table}[h]
    \centering
    \begin{tabular}{|c|c|c|c|c|c|}
        \hline
        Project & jGenProg & jKali & Nopol & ACS & Analysis \\ \hline\hline
        Chart   & 0/4      & 0/2   & 1/1   & 0/0 & 7/3      \\ \hline
      Closure    & --      & --   & --   & -- & 8/2      \\ \hline
        Lang    & 0/0      & 0/0   & 0/0   & 1/0 & 10/0     \\ \hline
        Math    & 2/1      & 1/1   & 0/0   & 3/0 & 8/2      \\ \hline
        Time    & 0/1      & 0/1   & 0/0   & 0/0 & 9/1      \\ \hline\hline
        Total   & 2/6      & 1/4   & 1/1   & 4/0 & 42/8     \\ \hline
    \end{tabular}
    \caption{Comparison our analysis result with existing automatic
      repair techniques on our dataset. 
      The results of the first three approaches come from
      \citep{durieux2015automatic} and the result of ACS comes from
      \citep{xiong2016precise}. The data of {\it Closure} is missing because
      the evaluations of other approaches do not include {\it
        Closure}. X/Y means that X defects are repaired correctly while
      incorrect patches are generated for other Y defects.
    }
    \label{tab:comparing}
\end{table}
}

\revise{We determine a patch as correct if the patch is either identical with or
semantically-equivalent to the standard patch. We say two patches
are semantically-equivalent only when we can transform one patch to
the other one by applying a series of semantics-preserving
transformations. For example, we show the two patches for
\textit{Chart-9} below. The first patch was generated 
by the developer in our case study 
and the second patch is the standard one provided by
Defects4J. The lines leading with ``+'' are added and those leading
with ``-'' are deleted. }{}

\revise{
In this example, the two patches are semantically equivalent because
in the first patch, the body of the two conditions are the same and
thus we can combine them into one. By further commutating the second
condition, the two patches are completely identical.
}{}


\finding{\revise{Under the same setting of automatic repair techniques, the
  human developer was able to fix defects with a recall of 84.0\%
and a precision of 89.4\%, indicating that automatic defect repair
techniques have much room to improve.}{
Up to 84\% defects can be correctly fixed in our analysis, 
indicating that most of the defects have a great potential
to be fixed under a weak test suite.
}}

\revise{
We further investigated why the developer generated five incorrect
patches, and we found out that in all the five cases, the tests in the
program do not provide enough information to reveal the full scope of
the defect. Without knowing the precise specifications of the programs, the
developer only provided incomplete patches. 
}{
Among the 8 defects classified as \emph{difficult to repair}, we
generated incorrect patches for 5 defects while generated correct
patches for 3 defects. We further investigated the 5 incorrect
patches, and found out that in all those cases, the tests in the
program do not provide enough information to reveal the full scope of
the defect. Without knowing the precise specifications of the programs, we
would generate incomplete patches based on only the test suite.
}
For example, a defect from \textit{Chart-10} is related to
\code{String} transformation. According to the failing test, character
\textcolor{blue}{\small $\backslash"$} in the input should be replaced
with \textcolor{blue}{\&quot;}. Based on this, \revise{the developer}{we} generated
the statement
\code{toolTipText=toolTipText.replaceAll} \code{("$\backslash\backslash\backslash$"","\&quot;");},
where variable \code{toolTipText} contains the input {string}. 
This patch passed all the tests.
However, compared with standard patch, we found that many other
special characters should be replaced besides \textcolor{blue}{\small
  $\backslash"$}, e.g., \textcolor{blue}{ \&} should be replaced with \textcolor{blue}{\&amp}. 
However, without knowing a complete list of characters to be replaced,
it is not possible to generate the correct patch.

\revise{}{Interesting, though the incorrect patch generated by
  automatic approaches often break the existing functionality of a
  program, we do not observe such behavior in the incorrect patches
  generated in our analysis. The incorrect patches are all due to incompleteness.}

\revise{
We then investigated why the developer failed to fix the three defects. 
}{
We further investigated why we could not generate a
patch for the three defects during our analysis. 
} 
The reason is similar: these defects require domain knowledge
either specific to the project or specific to a particular domain,
where \revise{the developer was}{an average developer may} not be familiar with. Among the three defects,
\textit{Math-2} is a defect about floating-point precision, where the standard
patch changes inaccurate expression into a mathematically equivalent
but more accurate expression. Fixing the defect requires the knowledge
of accurate arithmetic. \textit{Closure-4} and \textit{Time-6} are related to
the uses of the methods and classes in the project, where the buggy
code does not correctly interpret the semantics of called methods
or the preconditions of called methods are not properly satisfied.
Fixing the defect requires the knowledge of the project, especially
the preconditions and semantics of each method.
Lacking the domain knowledge, \revise{the human developer failed}{it is
  difficult for an average developer} to locate the
root cause of the three defects.



\finding{
Incomplete specifications or domain knowledge about programs may lead to
overfitting patches or falures of identifying the root causes of the defects,
indicating that \revise{lack of specifications may indeed block the repair of some
defects.}{weak test suites indeed impose challenges in defect repair,
even for human developers.}
%
}


\revise{
Finally, we analyze the time used by the developer. We classify the
defects into six categories based on their repair time: defects
repaired within five minutes, between ten and twenty minutes, half to
one hour, within two hours, within three hours, and those with longer time. Figure \ref{fig:time} shows the number of defects for each category. 
}{}

\revise{
\begin{figure}[h]
    \includegraphics[width=1.0\textwidth]{time}
    \caption{Distribution of repair time among the defects 
    }
    \label{fig:time}
\end{figure}
}{}

\revise{
From Figure~\ref{fig:time}, we can find that more than 92\% defects in our experiment dataset can be fixed by our developer within two hours, leaving the average time for fixing one defect is 47 minutes and the median is 20 minutes. 
}{}

\revise{
\finding{Some defects can be repaired in short time even for
  a developer unfamiliar with the project, while some defects require
a long time to repair.}
}{}

\subsection{RQ2: \revise{Methodology for Human-Repair}{Methodology for Repair}}
\label{subsec:rq2}
\revise{
In this subsection we analyze how the developer derived the patches
from the program and the tests. Following the usual design of
automatic repair process, we view the defect repair process as two
sub-processes: fault localization and patch generation, and try to
understand the manual repair process from the two aspects.
}{
In this subsection we summarize how we derived the patches from 
the program and the tests. Following the usual design of automatic repair
process, we view a repair process as two sub-processes: fault localization and patch generation. The first
sub-process is identifying the root cause of the failure, based on which
the second sub-process is generating a patch that can fix the failure.
We shall decompose the repair process from the two aspects.
}

In an abstract view, both the fault location and the patch generation
can be seen as locating a solution in a (possibly finite or infinite)
space of solutions. In fault localization, the space is the power set
of all statements and we try to locate one statement or a few statements
that is the root cause of the defect. In patch generation, the space
is all possible patches and we try to generate a patch that can fix the
current defect.

\revise{
Based on our current understanding after analyzing the repair records, we
view both the fault location and the patch generation as processes of
assigning confidence values to the elements in the spaces. When the
confidence value of one element is significantly higher than other
elements, and the difference exceeds a specific threshold, 
the developer claims that a solution is found.
}{
To understand how the defects can be repaired, we need to decompose the fault
localization process and the patch generation process used in our
analysis. However, how human debugs is an open problem in general and there lack
theories to support this decomposition. To provide useful guidance for
automatic repair techniques, here we assume a
model with strategies and try to derive strategies from our analysis.
Concretely, we 
view both the fault localization and patch generation
as processes of assigning confidence values to the elements in the spaces.
When the confidence of one element is significantly higher than other
elements and the difference exceeds a specific threshold, we would 
try to generate a patch and verify its correctness by running the test cases.
This process was iteratively proceeded until a patch passed all the test cases.
In each iteration, a strategy is applied to adjust the confidence values.
A strategy, when applied, either increases or decreases the
confidence values of some solutions in the space. A strategy is
usually associated with a precondition, which must be satisfied 
before applying the solution. During the analysis, we always need 
to simultaneously consider a large set of strategies and determine 
which of them can be applied.}

\revise{
In a more fine-grained level, the manual repair process can be seen as a
series of attempts to apply different strategies to the current
problem. A strategy, when applied, either increases or decreases the
confidence values of some solutions in the space. A strategy is
usually associated with a precondition, which must be satisfied before
applying the solution. A human developer seems to be able to
simultaneously consider a large set of strategies and determine which
of them can be applied.
}{
In a more fine-grained level, 
the analyzing process is a series of attempts 
to apply different strategies to the current problem.
A strategy, when applied, either increases or decreases the
confidence values of some solutions in the space. A strategy is
usually associated with a precondition, which must be satisfied 
before applying the solution. During the analysis, we always need 
to simultaneously consider a large set of strategies and determine 
which of them can be applied.}
For example, a simple strategy of fault localization is to exclude all
statements that are not executed during the failed test execution.
This is equivalent to (greatly) decrease the confidence values of all
solutions containing these statements. This strategy can be applied
only when there is an executable test that is failed by the program
(this precondition is always satisfied under the setting of automatic
defect repair). As another example strategy, if we observe a rare
statement that breaks usual code conventions, such as \code{if(a=1)}
rather than \code{if(a==1)}, we can increase the confidence value of
this statement during fault localization.

\revise{\finding{\reviseno{The manual repair process is a series
of implicit attempts to apply different strategies to increase
or decrease the confidence values of some solutions.
}{
In the analysis, we usually implicitly attempts to apply 
different strategies to increase or decrease the confidence 
values of some solutions based on the program runtime 
information and our domain knowledge.
}}}{}





\revise{
Under this view, the understanding of the manual repair process
becomes the understanding of the strategies. 
After analyzing all the manual repair records, 
we identified seven strategies for fault
localization and seven strategies for patch generation. 
}{
Under this view, to understand how the defects with weak test suites could 
be fixed, we try to decompose the repair processes for the 42 defects
into a set of strategies.}
Totally, we identified seven strategies for fault
localization and seven strategies for patch generation
based on our analysis.
A further observation on the strategies is that the distinction between fault
localization and patch generation is not always clear. A strategy can
contribute to both fault localization and patch generation. 
For example, the aforementioned strategy on code convention not only
gives us confidence on fault localization, but also gives us a
solution in patch generation.

\revise{\finding{The distinction between fault localization and patch
  generation is not always clear. A strategy can contribute to both.}}{}

In the follows we introduce the strategies for fault localization and
for patch generation, respectively. If a strategy
contributes to both sub-processes, we classify it based on its main contribution.

\subsubsection{Fault Localization Strategies}
The seven fault localization strategies are listed in Table \ref{tab:loc-strategy}.
The first column lists the strategy names, the second column briefly
describes how these strategies work, and the last column lists the
defects to which each strategy was applied during the \revise{manual repair}{analysis}
process.

\begin{table}[h]
    \centering
    \begin{tabular}{p{2.0cm}|p{6cm}p{2.5cm}}
           \hline
        \textbf{Strategy} & \textbf{Description} & \textbf{Defects} \\
        \hline\hline
        Excluding unexecuted statements    & 
        Exclude those statements not executed by failing test.   & 
        All defects. \\
        \hline
        Excluding unlikely candidates    & 
        \revise{}{Filter all non-related candidates based on their functionalities and complexities.}   & 
        \begin{tabular}[c]{@{}l@{}}\textit{Lang-1,2,4,7,9}\\ \textit{Math-5,10}\\ \textit{Chart-2}\\ \textit{Closure-9}\\ \textit{Time-1,4,10} \end{tabular} \\
        \hline
        Stack trace analysis     & 
        Locate faulty locations based on the stack trace information thrown by failing test cases.  & 
        \begin{tabular}[c]{@{}l@{}}\textit{Lang-1,5,6}\\ \textit{Math-1,3,4,8}\\ \textit{Chart-4,9}\\ \textit{Closure-2}\\ \textit{Time-2,5,7,8,10} \end{tabular} \\
        \hline
        Locating undesirable changes     & 
        \revise{Locate those statements that cause the difference between real output and the expected output of failing test.}{Locate those statements that generate the final faulty values of failing test cases.}   & 
        \begin{tabular}[c]{@{}l@{}}\textit{Lang-8}\\ \textit{Closure-1,3,5,7,8,10}\\ \textit{Time-3,9} \end{tabular} \\
        \hline
        Checking code conventions     & 
        Identify those code that obviously violate some programming principles based on previous programming experience.   & 
        \begin{tabular}[c]{@{}l@{}}\textit{Lang-6,8}\\ \textit{Chart-1,7,8} \end{tabular} \\
        \hline
        Predicate switching     & 
        Inverse condition statements to get expected output, the inversed condition statement is the error location.   & 
        \begin{tabular}[c]{@{}l@{}}\textit{Lang-3}\\ \textit{Chart-1,9}\\ \textit{Closure-10} \end{tabular} \\
        \hline
        Program understanding     & 
        \revise{Understand the logic of faulty program and then combine domain knowledge to locate error locations.}{Understand the logic of faulty program and the functionalities of relevant objects and methods.}  & 
        \begin{tabular}[c]{@{}l@{}}\textit{Lang-10}\\ \textit{Math-6,9}\\ \textit{Chart-3}\\ \textit{Closure-9}\\ \textit{Time-3,9} \end{tabular} \\
        \hline
    \end{tabular}
\caption{Strategies applied \revise{by our developer to locate faulty method.}{to locate faulty method in our analysis.}}
\label{tab:loc-strategy}
\end{table}

\newcounter{strategy}
\newcommand{\strategy}[1]{\refstepcounter{strategy}
  \smallskip
  \textit{\textbf{Strategy \arabic{strategy}:} {#1}}.
}

\strategy{Excluding unexecuted statements\label{stg:execution}}

This strategy is very simple: when a statement is not executed during
the execution of the failed test, it cannot be the root cause. 
\revise{Then human developer implicitly applies this strategy when he starts to
debug a defect in a debugger, and the debugger will guide him to only
examine the executed statements. This strategy applies to almost all defects.
}{
This strategy is implicitly applied when we try to find the root cause of a defect.
Actually, this strategy can be applied to almost all defects.}

\strategy{Excluding unlikely candidates}\label{stg:unlikely}

Given a list of possible candidates of root causes, \revise{the developer could}{we could}
examine them one by one, and exclude those that are unlikely to
contain defects. Though technically this strategy can be applied to
different granularities, 
\revise{we found that in this case study the developer applies it only 
on the method level. That is, given a list
of methods invoked during the failed test execution, the developer may
examine them one by one and exclude unlikely ones. The developer
mainly uses the following three criteria to determine the safety of a method.
}{
we found that applying it on the method level was effective in our analysis.
That is, given a list of methods invoked during the failed test execution, 
we will examine them one by one and exclude unlikely ones. We
found that the following three criteria are effective on our dataset.
}
\begin{itemize}
\item When a method is a library function, it is unlikely to contain
  defect.
\item In Java, because the lack of default parameter or the use of
  polymorphism, it is often the case that a method is just a wrapper
  of another method, and the purpose is only to pass a default
  argument or to adapt to an interface. When a method is a simple
  wrapper method, this method is unlikely to contain defect.
\item When a method is the test itself, it is unlikely to contain
  defects. 
\end{itemize}

Note that technically the methods excluded by this strategy still have
the possibility to contain defects, but their probability is
significantly smaller than other methods.

\begin{figure}[ht]
  \centering
   \includegraphics[width=1.0\textwidth]{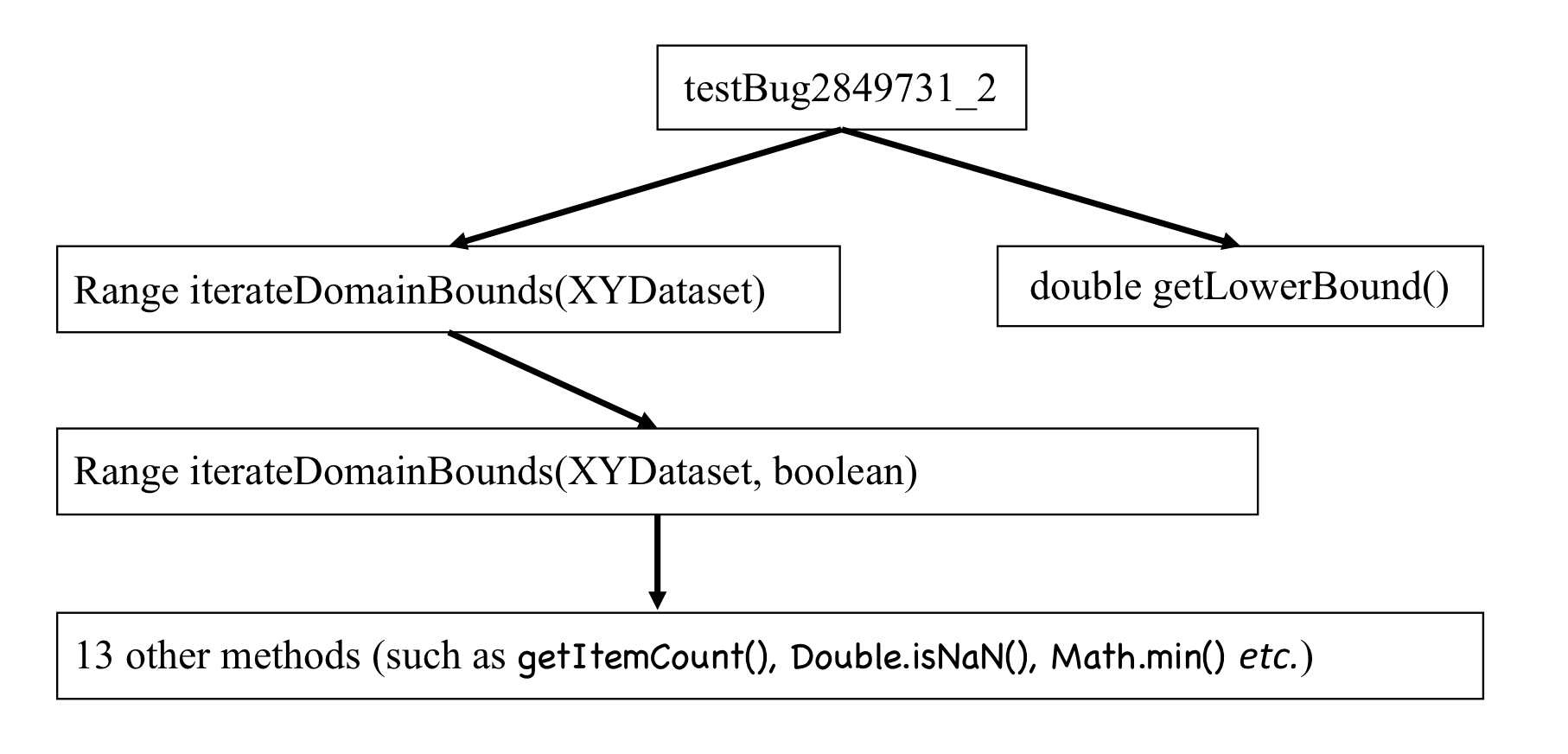}
  \caption{The call graph of {\it Chart-2} \label{fig:chart2}}
\end{figure}

Excluding unlikely candidates was a very effective strategy in our
\revise{case study}{analysis}, as \revise{the developer}{we} could localize the faulty method using
only this strategy and strategy~\ref{stg:execution}. For example,
Figure~\ref{fig:chart2} shows the call graph of defect
\textit{Chart-2}. As we can see from the figure, the test calls in
total 16 methods directly or indirectly. Based on
strategy~\ref{stg:execution}, all of them may be faulty. However, many
of them are library methods, such as 
\code{Double.isNaN} and \code{Math.min}. Many other methods are
simple wrapper method, such as \code{iterateDomainBounds(XYDataset)},
whose implementation code is listed below.
\begin{lstlisting}[style=java,numbers=none]
    public static Range iterateDomainBounds(XYDataset dataset) {
        return iterateDomainBounds(dataset, true);
    }
\end{lstlisting}
Since this method is very simple, it is unlikely to contain defect.

After we have excluded all these methods, the only remaining method is
\code{iterateDomainBounds(XYDataset,boolean)}, which turned out to be the faulty method.
\revise{}{Apparently, during this process, we need not to know the specifications 
about the program and even need not to understand the full functionality of the relevant methods.}

\strategy{Stack trace analysis}\label{stg:stack}

When an uncaught exception is triggered, the program crashes and the
stack trace information is printed. A stack trace lists a sequence of
locations in the program where a method is called but is not returned
before the point of crash. Usually the root cause of the
fault is close to the locations listed in the stack trace. 
That is, the confidence of the statements near the locations in the
stack trace increases while the confidence of other statements decreases.
In our \revise{case study}{analysis}, 15 defects were localized with the contribution of
this strategy. Further combined with other strategies, we can often
locate the root cause of the defect.

\begin{figure}[h]
    \includegraphics[width=1.0\textwidth]{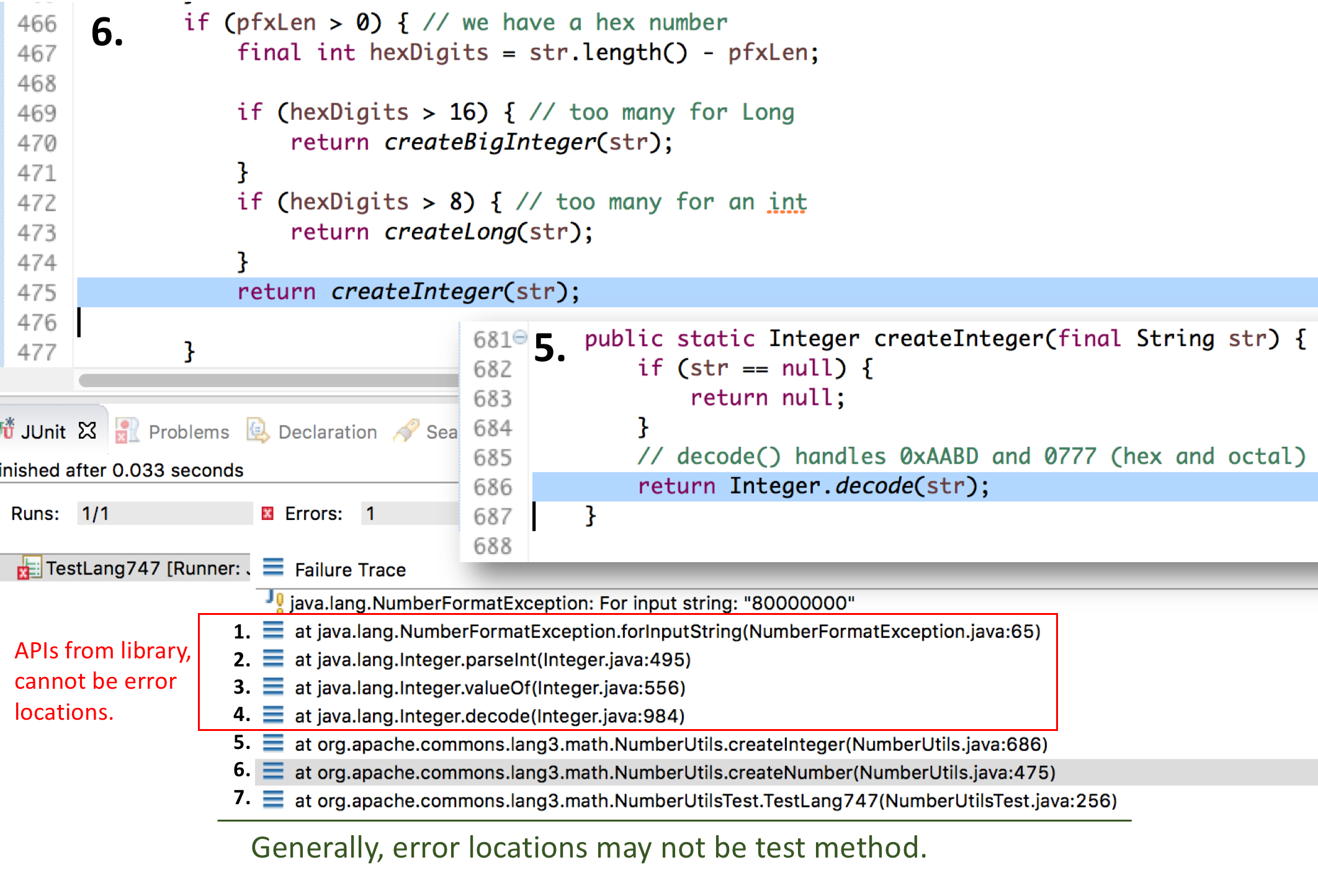}
    \caption{Screenshot for the stack trace of \code{java.lang.NumberFormatException}, which comes from defect \textit{Lang-1}. In this figure, line 469 and 472 are buggy conditions while line 475 is reported via failure trace.}
    \label{fig:loc-stacktrace}
\end{figure}

For example, Figure \ref{fig:loc-stacktrace} is the
screenshot when \revise{debugging}{analyzing}
the defect \textit{Lang-1}, which throws an uncaught exception,
\code{NumberFormatException}. 
The stack trace lists seven locations in the program. 
Then we can filter the locations based on strategy~\ref{stg:unlikely}. 
Among them, the first four locations are related to the library
APIs(\code{java.lang}), the last error
location is in the test method, and the fifth location is a simple
wrapper method (showed in the upper part of
Figure~\ref{fig:loc-stacktrace}). After filtering all of them, the
only possible location is the sixth.

\strategy{Locating undesirable changes}\label{stg:undesired}

A failed test execution produces an output that is different from the
desired output. However, sometimes the desired output has already been
constructed during the test execution, but the execution of some
statements, $S$, turn it into an undesirable one. In such cases, $S$
or the statements that $S$ control dependent on are likely to be
faulty. Note that the latter should be included because they are the
reason why $S$ is executed.



\begin{figure}[t]
    \includegraphics[width=1.0\textwidth]{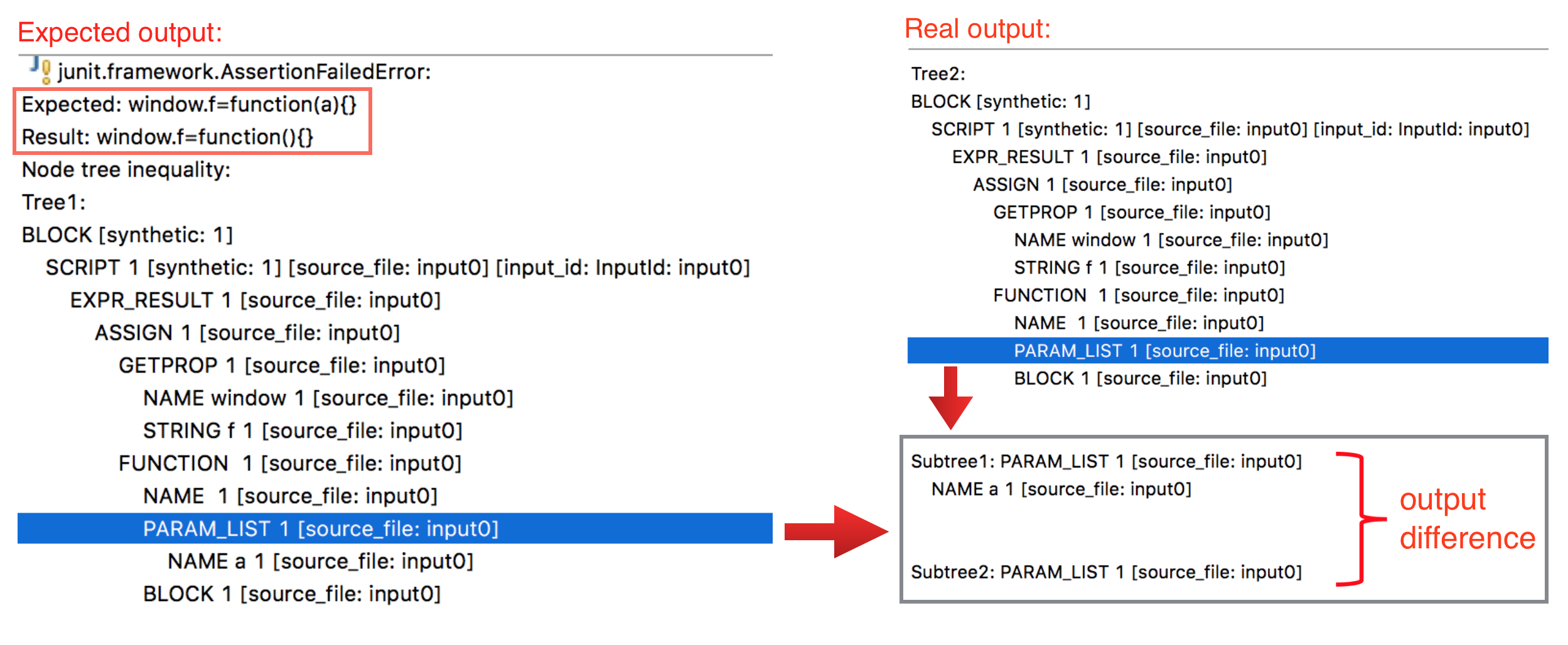}
    \caption{Screenshot for the difference between expected and real output, which comes from defect \textit{Closure-1}. The test input is the JavaScript source code \code{window.f = function(a) \{\};}. After optimization, the parameter \code{a} is deleted and the differences on abstract syntax tree are marked by blue box.}
    \label{fig:loc-diff}
\end{figure}

In our \revise{case study}{analysis}, those cases frequently occur when testing the
optimization component in the closure project. In a typical such
defect, the test input is a program that should not be optimized, and
the execution output is an undesirably optimized program, which is not
semantically-equivalent to the original program. In such a case, we
can increase the confidence values of the statements that make such an
undesirable change. For example, 
Figure \ref{fig:loc-diff} shows the output information
for a failing test from \textit{Closure-1}. The test input and the
expected output are both \code{window.f=function(a)\{\}} while the
actual output does not contain the parameter \code{a}. By examining
the execution trace, \revise{the developer}{we} noticed that \code{a} was removed
in a method call \code{removeChild}, so either this method or the
statements leading to the call of this method might be faulty.
\revise{}{This process is really a mechanical tracking of the variables.}

\strategy{Checking code conventions}\label{stg:convention}

Though in principle the language construct can be combined in any way
to form a program, in practice people would only use a small subset
of combinations. Basically, a code convention defines a
constraint on combining the language construct, and a piece of code
violating the constraint is likely to be faulty. A typical code
convention, as mentioned before, is that assignment is unlikely to be
used in an ``if'' condition, and thus statement like \code{if(a=0)} is
likely to be faulty\footnote{Please note that this code convention is
  useful for C but not on Java, as \code{if(a=0)} will cause type
  error in Java. We cite this example just for illustration, and this
  is not a convention we discovered.}. 
\revise{
We found that
the developer has the knowledge of a rich set of such code conventions
and implicitly check these conventions as he debugs along.
}{
We found that the violation of code convention is usually an
indication of fault and is useful in fault localization.
}


For example, the following piece of code shows the root cause of
\textit{Lang-6}. In this piece of code, there is a loop, however, the
body of the loop is kept accumulating an invariant value. Especially, this
value is obtained from a sequence by calling \code{codePointAt}. This
piece of code violates common code conventions and is likely to be faulty.

\begin{lstlisting}[style=java,numbers=none]
    for (int pt = 0; pt < consumed; pt++) {
        pos += Character.charCount(Character.codePointAt(input, pos));
    }
\end{lstlisting}



\strategy{Predicate switching}\label{stg:switching}

This strategy is very similar to the automatic fault localization
technique with the same name~\citep{zhang2006locating}. In some cases, if we inverse the result of an ``if''
condition and force the execution to switch to the other branch, the
failed test could pass. When such a case is observed, we may consider
the ``if'' condition may be faulty and increase
its confidence value.



For example, the following defect comes from \textit{Lang-3}. According to the failed test, when the test input is 3.40282354e+38, a \code{Double} number is expected while the following code returned a \code{Float} number. 
\revise{
    The developer noticed that if the first \code{if} condition was
    \code{false}, the next condition would be executed and might return
    the expected \code{Double} number to pass the failing test. Therefore,
    our developer suspected the first condition was wrong. 
}{
However, if we inverse the value of the first \code{if} condition to be false, the next condition would be executed and might return the expected \code{Double} number to pass the failing test. Therefore, we can increase the confidence of
the first \code{if} condition to be faulty.
}



\begin{lstlisting}[style=java, numbers=none]
   try {
     final Float f = createFloat(str);
     if (!(f.isInfinite()||(f.floatValue()==0.0F&&!allZeros))){
         return f;
     }
   } catch (final NumberFormatException nfe) {
   // ignore the bad number
   }
   try {
     final Double d = createDouble(str);
     if(!(d.isInfinite()||(d.doubleValue()==0.0D&&!allZeros))){
         return d;
     }
   } catch (final NumberFormatException nfe) {
   // ignore the bad number
   }
   return createBigDecimal(str);
\end{lstlisting}

\revise{Please note that when \reviseno{the human developer applies}{we apply} this strategy, it is
not a thorough and systematic examination of all ``if'' conditions as
in the automatic technique~\citep{zhang2006locating}. Basically,
\reviseno{when the developer debugged along}{in the analysis}, \reviseno{he}{we} would quickly glance at the code nearby
and judge whether inversing the condition can inverse the testing
result. If the condition is too far away or the judgment is too
difficult, \reviseno{he would simply give up this strategy.}{we would simply give up this strategy. Even thought, it was effective for locating 4 defects
  in our analysis.}}{}

\strategy{Program understanding}\label{stg:understanding}

The strategies we have seen so far can be applied without a full
understanding \reviseno{}{and specifications }of the program, and many faults can be localized by only
using these strategies. However, not all faults can be localized in
this way and a certain amount of program understanding is required.

Program understanding is a complex process and \revise{we were not able to obtain a
full understanding \reviseno{by analyzing the repair records.}{in the analysis.} However, some rough
understanding could be described. Roughly, \reviseno{the developer
  infers}{we often infer}}{here we try to describe it in terms of the general
logic reasoning process. Given a faulty program, we try to infer} likely
constraints on program behavior from different sources, and checks
consistency between them. If constraints inferred from different
sources are inconsistent, the related source code is likely to be faulty. On the
other hand, if constraints inferred from different sources are
consistent, the related source code is unlikely to be faulty. Typical sources
include the follows.
\begin{itemize}
\item Implementation Code. By interpreting the semantics of the source
  code, we can infer constraints on how the source transforms one
  state into another state.
\item Test Executions. Basically, each test gives a constraint on the
  desired output for each test input.
\item Identifier Names. \revise{The developer often infers}{We often try to infer} likely constraint from
  the names of the identifiers. For example, a method named ``remove''
  should reduce the number of items in some container. A variable
  named ``max'' should contain the maximum element in some container.
\item Comments. Sometimes the comments describe the intended behavior
  of a piece of code, and constraints could be inferred from the comments.
\end{itemize}

To understand how this strategy works, let us consider the defect {\it
Closure-1} which we have seen in strategy~\ref{stg:undesired}. 
\revise{
    Using strategy~\ref{stg:undesired}, we have localized this defect to method
    \code{removeChild} and its callers, and we know that the removal is
    undesired. From the name of \code{removeChild}, we can infer a
    constraint that this method should remove an item. Since this semantics is
    consistent with its implementation code, we know the removal within
    this method is desired, and the fault must be in the methods calling
    \code{removeChild}. In other words, \code{removeChild} should not be called.
}{
Using strategy~\ref{stg:undesired}, we can isolate the defect to method
\code{removeChild} and its callers, and we know the removal is undesired.
However, from the name of \code{removeChild}, we can infer a constraint
that this method should remove an item. Since this semantics is
consistent with its implementation code, we know the removal within
this method is desired. Therefore, the fault should be in the methods calling
\code{removeChild}. In other words, \code{removeChild} should not be called.}

Another example is \textit{Chart-3}. The patch of this defect is shown
below.
\begin{lstlisting}[style=java,numbers=none]
    TimeSeries copy = (TimeSeries) super.clone();
    //maxY(minY) saves the maximum(minimum) value in data
 +  copy.maxY = Double.NaN;
 +  copy.minY = Double.NaN;
    //data is reset but maxY and minY are not
    copy.data = new java.util.ArrayList();
\end{lstlisting}
The \code{TimeSeries} class contains three fields: \code{maxY},
\code{minY}, and \code{data}. From the identifier names we can infer
that \code{maxY} and \code{minY} possibly store some maximum/minimum
values calculated from \code{data}. However, the original code updates
\code{copy.data} but does not update \code{maxY} and \code{minY},
which violates this constraint and thus the code is possibly faulty.
\revise{}{Moreover, the failing test is related to the fields of \code{maxY} 
and \code{minY}, which increases the confidence that the code is faulty.}

\subsubsection{Patch Generation Strategies}

Table \ref{tab:fix-pattern} shows the seven strategies we summarized
on patch generation. Similarly to Table~\ref{tab:loc-strategy}, the
first column is the identification for each strategy, the second
column briefly describes the strategy, and the last column lists the
defects to which the strategy was applied.

\begin{table}[h]
    \centering
    \begin{tabular}{p{2cm}|p{6cm}p{2.5cm}}
        \hline
        \textbf{Strategy}  & \textbf{Description}      & \textbf{Defects}       \\
        \hline\hline
        Add NullPointer checker & 
        \revise{Add NullPointer checker to avoid NullPointerException}{Add NullPointer checker before using the object to avoid \code{NullPointerException}} &
        \begin{tabular}[c]{@{}l@{}}\textit{Math-4}\\ \textit{Chart-4}\\ \textit{Closure-2}\end{tabular}  \\
        \hline
        Return expected output   &
        Return the expected value according to the assertions. & 
        \begin{tabular}[c]{@{}l@{}}\textit{Lang-2,7,9}\\ \textit{Math-1,3,5,10}\\ \textit{Time-1,3}\end{tabular}  \\
        \hline
        Replace an identifier with a similar one  & 
        \revise{Replace an identifier with another one that has the similar name.}{Replace an identifier with another one that has the similar name and same type in the scope.}  &
        \begin{tabular}[c]{@{}l@{}}\textit{Lang-6,8}\\ \textit{Chart-7,8}\end{tabular}   \\
        \hline
        Compare test executions. &
        Generate patches by comparing the failed tests with those passed tests with similar test inputs. &
        \begin{tabular}[c]{@{}l@{}}\textit{Lang-2,5}\\ \textit{Math-1} \end{tabular}   \\
        \hline
        Interpret comments  &
        Generate patches by directly interpreting comments written in natural language.   &
        \begin{tabular}[c]{@{}l@{}}\textit{Math-9}\\ \textit{Closure-1,5,7,9}\\ \textit{Time-8,9}\end{tabular} \\
        \hline
        Imitate similar code element  &
        Imitate the code that is near the error location and has similar structure. &
        \begin{tabular}[c]{@{}l@{}}\textit{Lang-4,5}\\ \textit{Math-6,8}\\ \textit{Chart-1,2,7,9}\\ \textit{Closure-3,8,10}\\ \textit{Time-5,7,10}\end{tabular} \\
        \hline
        Fix by program understanding  &
        \revise{Generate patches based on program understanding.}{Generate patches by understanding the functionality of program.}    &
        \begin{tabular}[c]{@{}l@{}}\textit{Lang-1,3,9,10}\\ \textit{Math-6,9}\\ \textit{Chart-2,3}\\ \textit{Closure-3,8}\\ \textit{Time-1,2,4,10}\end{tabular}    \\
        \hline
    \end{tabular}
    \caption{Strategies used \revise{by our developer when generating patches.}{to generate patches in our analysis.}}
    \label{tab:fix-pattern}
\end{table}

\strategy{Add NullPointer checker}\label{stg:nullchecker}

This strategy was usually used \revise{by the developer}{in our analysis} when a test failed
because of \code{NullPointerException}. A typical way to fix such a
defect is to surround the statement and all following dependent
statements with a guarded condition \code{x != null}, where \code{x}
is the variable causing this exception.


For example, the following code is the patch for \textit{Chart-4}.

\begin{lstlisting}[style=java, numbers=left]
    XYItemRenderer r = getRendererForDataset(d);
    if (isDomainAxis) {
        if (r != null) {
            ...
        }
        ...
    }
    ...
 +  if(r != null){
         Collection c = r.getAnnotations();
         Iterator i = c.iterator();
         while (i.hasNext()) {
             XYAnnotation a = (XYAnnotation) i.next();
             if (a instanceof XYAnnotationBoundsInfo) {
                 includedAnnotations.add(a);
             }
         }
 +  }
\end{lstlisting}
In this patch, an exception is thrown at line~10. The patch adds an
``if'' statement to surround line~10 and all following
statements that depend on line~10.

Please note that though null checker is often added for \code{NullPointerException},
the strategy alone usually cannot decide a patch. In this case, we may
also change the method \code{getRendererForDataset} so as not to
return \code{null}. \revise{The developer comes}{We come} to this patch by further
considering two facts: (1) applying this patch makes all tests pass,
(2) there is also a checker for \code{r} at line~3, indicating that
returning \code{null} is a valid behavior of
\code{getRendererForDataset}. We use
strategy~\ref{stg:patchunderstanding} to summarize these
consideration, which will be explained later.



\strategy{Return expected output}\label{stg:expected}

When programming, we often encounter boundary cases that should be
considered separately from the main programming logic, and such
boundary cases are easily neglected by developers. A boundary case is
typically handled by a statement \code{if(c) return v}, where \code{v}
is the expected result and \code{c} is a condition to capture the
boundary case.

As a result, if the failed test execution is a boundary case, we
may consider patches that handle the boundary case using the
above form. For example, the following code snippet is a failing test from \textit{Math-3}.


\begin{lstlisting}[style=java, numbers=none]
    public void testLinearCombinationWithSingleElementArray() {
        final double[] a = { 1.23456789 };
        final double[] b = { 98765432.1 };
        Assert.assertEquals(a[0] * b[0], 
                    MathArrays.linearCombination(a, b), 0d);
    }
\end{lstlisting}

This test calls method \code{linearCombination} with two arrays of
length one. 
\revise{
    Based on \revise{his}{previous} programing experience, \revise{our developer realized}{we can identify} 
    that an array of length one is a typical boundary case. Therefore, \revise{he}{we}
    generated a patch that inserts statement \code{if(len == 1){return
            a[0] * b[0];}} into the method \code{linearCombination}. 
}{
If we can identify that an array of length one is a boundary case, based on
above fixing form and the test case, we can easily come to the fix as
inserting statement \code{if(len == 1){return a[0] * b[0];}} into the method
\code{linearCombination}.}
Here variable \code{len} is used because
\code{linearCombination} requires the two input arrays to have the
same length and uses the variable \code{len} to store their length. The
expression \code{a[0]*b[0]} is just the expected result.

Note that the use of this strategy heavily depends on the developer's
experience to determine boundary cases. Otherwise the generated patch
may overfit to the current test suite.

\strategy{Replace an identifier with a similar one}\label{stg:replacevar}

When the names of two identifiers are similar, developers may confuse
the two identifiers. As a result, a possible patch is to replace an
identifier with another one whose name is similar. Of course,
this strategy alone can hardly determine a patch, but this strategy
can be used together with other strategies to increase the confidence
of some patches.

For example, in defect {\it Lang-6} we have seen in
strategy~\ref{stg:convention}, we can observe that two variables,
\code{pos} and \code{pt}, have very similar names. In fact, if we
replace the last occurrence of \code{pos} with \code{pt}, we find that
the piece of code no longer violates the code convention. Furthermore,
rerunning all tests could reveal that this patch passes all the tests.
Putting all together, we gain enough confidence on the correctness of
this patch.

\strategy{Compare test executions}\label{stg:compExecution}

It is common that more than one test case exists to test a specific
method, and only one of them fails. By comparing the passed tests and
the failed tests, we can often obtain useful information on patch
generation.

For example, the following code is the patch for \textit{Lang-2}.

\begin{lstlisting}[style=java, numbers=none]
    public static Locale toLocale(final String str) {
        if (str == null) {
            return null;
        }
 +     if(str.contains("#")){
 +          throw new IllegalArgumentException("Invalid locale format: " + str);
 +     }
        ...
    }
\end{lstlisting}

This method is used to transform a \code{String} to a \code{Locale}
object. The failed test has an input of
``ja$\_$JP$\_$JP$\_\sharp$u-ca-japanese'' and expects an exception
\code{IllegalArgumentException}. While throwing an exception seems to
be a boundary case which we can handle using
strategy~\ref{stg:expected}, it is not clear what condition we can use
to capture the boundary case. 
\revise{
By comparing the failed test with other passed tests, 
the developer noticed that all passed tests do not
contain the character $\sharp$. Thus, containing character $\sharp$ is
probably a boundary case and the above patch can be generated.
}{
However, if we compare the failed test case with all the passed test cases,
we can notice that all passed test cases do not contain the character $\sharp$,
which provides it a high confidence that containing the character $\sharp$ is
probably a boundary case. Therefore, putting them together,
the above patch can be generated.
}


\strategy{Interpret comments}\label{stg:comment}

Program source code may contain comments explaining properties of the
program, such as functionality, precondition, and etc. In particular,
Java programs often come with Javadoc annotations explaining the
method, the parameters, the return value, and exceptions that might be
thrown. By interpreting these comments, \revise{the developer}{we} can often gain
confidence on some patches.


For example, the following method was used to create a
\code{DateTimeZone} object based on the given hour and minute, which comes from \textit{Time-9}. The
failed test expects an 
\code{IllegalArgumentException} to be thrown at the input of 24 and 0.
Again, this is a boundary case where strategy~\ref{stg:expected} can
be applied. However, we still do not know what condition should be
used to capture this boundary case. By reading the Javadoc annotation,
we can know that the input should be in the range of $-23\sim23$, and
\reviseno{we can generate the following patch.}{
    the following patch is straightforward.}

\begin{lstlisting}[style=java, numbers=none]
   /*@param hoursOffset
    *       the offset in hours from UTC, from -23 to +23
    */
    public static DateTimeZone forOffsetHoursMinutes(
         int hoursOffset, int minutesOffset) 
         throws IllegalArgumentException{
  +      if(hoursOffset < -23 || hoursOffset > 23){
  +          throw new IllegalArgumentException();
  +      }
        ...
    }
\end{lstlisting}

\strategy{Imitate similar code element}\label{stg:imitatecode}

In general, programs with similar functions often have similar
structures. When similar code pieces exist near the buggy code, we can
generate patch by imitating the similar code. This strategy is often
useful when we found the program fails to handle some cases, but we do
not know how to handle these cases without the full specification.
However, if we can find code pieces handling similar cases, we can
imitate these code pieces.

For example, the following patch comes from {\it Chart-9}, which we have seen in Section~\ref{subsec:rq1}.
According to the failing test, when the \code{startIndex} is greater than \code{endIndex}, no exception should be thrown, 
which can lead us to generate the condition statement \code{if(startIndex $>$ endIndex)}.
However, we do not know what object should be returned in the body of \code{if} condition.
By reading the code nearby, we find that the first \code{if} condition is used to handle a similar case, so we can generate the 
following patch by imitating the first \code{if} condition.

\begin{lstlisting}[style=java, numbers=none]
    if(endIndex < 0){
        emptyRange = true;
    }
  + if(startIndex > endIndex){
  +    emptyRange = true;
  + }
    if (emptyRange) {
        TimeSeries copy = (TimeSeries) super.clone();
        copy.data = new java.util.ArrayList();
        return copy;
    }
\end{lstlisting}

\strategy{Fix by program understanding}\label{stg:patchunderstanding}

Similar to fault localization, this strategy is placed to capture the
case \revise{where the developer generates the patch with his understanding of
the program.}{where we generate the patch by understanding the functionality of the program.}
The process is similar to the fault localization case,
but the potential patches become another source for generating
constraints. 
\revise{If the developer}{If we} found the constraints generated from a
patch are consistent with all other constraints, \revise{he}{we} would increase the
confidence value of the patch.

Similar to fault localization, we still lack a full understanding of
the program understanding process, and future work is needed to
further understand this process.

\subsection{RQ3: Inspiration from Analysis}
\label{subsec:rq3}

\revise{
To understand to what extent the manual repair process can be
automated, we examine the strategies defined in
Section~\ref{subsec:rq2} and check how much they can be automated. Our
observation is that most of the strategies are simple heuristic rules
that do not require deep semantic analysis or full understanding of
the program, indicating a high possibility to automate them. 
Many strategies perform only mechanical operations and can be
easily automated. 
For example, \textit{Stack trace analysis} and \textit{Locating
  undesirable changes} performs only mechanical operations. Some
strategies require human experience, but such experience has a high
potential to be summarized as heuristic rules. For example,
\textit{Excluding unlikely candidates} relies on a few heuristic rules
to determine whether a method may be faulty.
In fact, only the last strategy in each category,
strategy~\ref{stg:understanding} and
strategy~\ref{stg:patchunderstanding}, require program understanding.}{}

\revise{}{
    To understand how much can current program repair techniques be improved
    based on our analysis, we examine the strategies defined in
    Section~\ref{subsec:rq2} and check how much possible they can be automated. Our
    observation is that most of the strategies are simple heuristic rules
    that do not require deep semantic analysis or full understanding of
    the program, indicating a high possibility to automate them. 
    Many strategies perform only mechanical operations and can be
    easily automated. 
    For example, \textit{Stack trace analysis} and \textit{Locating
        undesirable changes} performs only mechanical operations just as 
    explained in Figure~\ref{fig:loc-stacktrace} and \ref{fig:loc-diff}. Some
    strategies require human experience, but such experience has a high
    potential to be summarized as heuristic rules. For example,
    \textit{Excluding unlikely candidates} relies on a few heuristic rules
    to determine whether a method may be faulty. Some simple strategies, 
    such as excluding library functions, used in our analysis have been listed 
    in strategy~\ref{stg:unlikely}, which can be easily expanded by developers.
    In fact, only the last strategy in each category,
    strategy~\ref{stg:understanding} and
    strategy~\ref{stg:patchunderstanding}, require full program understanding.}

\revise{This result is further confirmed by an analysis of the repair records,
where we found that the developer did not obtain a full understanding
of the method being debugged for at least 33 of the defects.
}{
  As a simple classification, we can consider the two program
  understanding strategies as difficult to automate, and the other
  strategies as easy to automate. Then we found that there are at
  least 25 defects that can be fixed by only easy-to-automate strategies, consisting of 50\% of all defects. As shown in
  Table~\ref{tab:loc-strategy}, existing program repair approaches can fix much fewer
  defects. This indicates potential rooms for improving current
  techniques. 
}

\finding{Many strategies are simple heuristic rules that do not
  require deep analysis nor full understanding of the defects,
  indicating possibility of automating these strategies \revise{}{to improve current automatic repair techniques}.}

We further observe that, many of the strategies have already been
considered in automatic program repair techniques. 
\revise{}{However, these techniques often have weak performance than
  the strategies we considered.}
In the following we try to analyze the most
related work within our knowledge for each strategy
\revise{.}{ and identify the concrete places where current techniques
  can be improved.}
Please note that this analysis is not an attempt to thoroughly summarize 
the related work in fault localization and patch generation, and
readers are redirected to recent
surveys~\citep{Wong2016A,monperrus2015automatic} for such a
summarization.

\def\tech{\textbf{Related Work:  }}
\def\improve#1{\textbf{Improvements:  {\it #1}}}
\def\stg#1{\smallskip \noindent $\blacktriangleright$ #1}

\stg{Excluding unexecuted statements}
    \begin{itemize}
        \item \tech This strategies is almost adopted in any fault localization approaches.
        \item \improve{} None.
    \end{itemize}
\stg{Excluding unlikely candidates}
    \begin{itemize}
        \item \tech This strategy relies on the
        features of the candidate methods to exclude the unlikely ones. A
        related approach is fault prediction~\citep{hall2012systematic},
        which predicates the probabilities of different software components
        to contain defects based on features of the software components. \revise{Our result indicates that fault prediction can be potentially integrated into fault localization to raise the accuracy of fault localization.}{}
        \item \improve{\revise{}{Incorporate dynamic information of test failure.}}\\
        \revise{}{Current fault localization techniques judge whether a given method is correct or
        not only based on the static features of the elements but without
        considering the relationship between the current failure and
        the method, which may cause incorrect judgment.
        For example, the following code is the patch for defect {\it Chart-6}.
        Simply according to the static feature of the code, it is very likely
        to regard it as correct by current fault prediction techniques because of its simpleness.
        However, by collecting the states of input object we can find
        that all the input objects have the same type of \code{ShapeList},
        which is consistent with the type required by this method.
        On the contrary, the \code{super.equals(obj)} calls the method
        \code{equals} in class \code{AbstractObjectList}, which requires type 
        of \code{AbstractObjectList}. Putting all these together, we suspect
        the following method is faulty. From this example, we can see that
        sometimes static features are not enough to decide the correctness of
        a method while dynamic information may provide helpful guidance.}
    \end{itemize}
\begin{lstlisting}[style=java,numbers=none]
    public boolean equals(Object obj){
      if(obj==this){
        return true;
      }
      if(!(obj instanceof ShapeList)){
        return false;
      }
  +   ShapeList that=(ShapeList)obj;
  +   int listSize=size();
  +   for(int i=0;i<listSize;i++){
  +     if(!ShapeUtilities.equal((Shape)get(i),
  +                     (Shape)that.get(i))){
  +       return false;
  +     }
  +   }
  +   return true;
  -   return super.equals(obj);
     }
\end{lstlisting}
\stg{Stack trace analysis}
    \begin{itemize}
        \item \tech Stack trace analysis has been
        adopted by many fault localization approaches. For example, \cite{Wu:2014:CLC:2610384.2610386} propose a fault localization
        approach mainly based on stack trace information. \cite{wong2014boosting} propose to combine stack trace analysis
        with bug reports to enhance the accuracy of fault localization.
        \item \improve{} None.
    \end{itemize}
\stg{Locating undesirable changes}
    \begin{itemize}
        \item \tech Within our knowledge, this
        strategy is not directly adopted by existing fault localization
        approaches. A loosely related approach\revise{}{, delta debugging,} is proposed by~\cite{cleve2005locating} to locate the transitions that cause
        the fault. \revise{Delta} {However, delta} debugging requires \revise{a mechanism to determine the
        existence of faults for a test input, which does not apply to the
        setting of automatic program repair.}{(1) a mechanism to
        determine the test result and (2) a comparable passed test,
        which do not apply to the bugs solved by this strategy in our analysis.}
        \item \improve{\revise{}{Correctly identify undesirable changes.}}\\
        \revise{}{
          To overcome the problem, we need to introduce a new
          technique that could identify undesirable changes in a test
          execution. A possible way is to define a partial order
          between states to measure how close to the desirable state
          the current state is, where a standard test execution should
          only make the state more close to the desirable state rather
          than make it further.
        }
    \end{itemize}
\stg{Checking code conventions}
    \begin{itemize}
        \item \tech Static bug detection, such as
        FindBugs~\citep{ayewah2008using}, checks the
        conventions in the code to determine potential bugs. 
        \item \improve{Incorporate dynamic information of test failure.}\\
          \revise{Our result indicates that static bug detection could potentially be combined
            with fault localization approaches to better localize defects.
        }{
            Static bug detection faces the same problem with the strategy of {\it Excluding unlikely candidates}, where
            only considering the static information is not sufficient. For instance, considering the same example in strategy~\ref{stg:convention},
            we regard this code as a convention by combining multiple factors.
            First, the loop variable, \code{pt}, is not used in the loop body and second,
            the current failure is caused by the variable \code{pos} which is very similar to the variable \code{pt} with regard to their names and types.
            Finally, the value of \code{pt} is restricted by the length of the string \code{input} while \code{pos} is not.
            Therefore, if the variable \code{pos} is replaced by \code{pt} in the function call, the \code{IndexOutOfBoundException}
            can be avoid. From all above, we have enough confidence to say the code snippet is a code convention.
            To conclude, correctly checking code conventions not only needs to know the common convention
            patterns but also needs to combine the failure information.
        }
    \end{itemize}
\stg{Predicate switching}
    \begin{itemize}
        \item \tech As discussed before, this strategy is
        very similar to the predicate switching approach proposed by~\cite{zhang2006locating}.
        \revise{}{
            In fact, predict switching is even more powerful than that used in our analysis 
            because of computer's superb computation ability.
        }
        \item \improve{} None.
    \end{itemize}
\stg{Add NullPointer checker}
    \begin{itemize}
        \item \tech This strategy is similar to a
        template used in the repair approach
        PAR~\citep{tao2014automatically}, which applies a set of templates to
        the localized statement to
        generate patches.
        \item \improve{\revise{}{Correctly identify the location of the
            NullPointer checker.}}\\
          \revise{}{As discussed before, there are often more than one places to
          add the NullPointer checker, and identifying the correct
          location is the key for avoiding incorrect patches. In our
          repair process, different strategies are combined together
          to decide the correct location. This ability should be added
          to automatic program repair techniques.}           
    \end{itemize}
\stg{Return expected output}
    \begin{itemize}
        \item \tech This strategy is similar to a
        template used in ACS~\citep{xiong2016precise}.
        \item \improve{\revise{}{Correctly identify boundary cases.}}\\
         \revise{which relies on a precise
            condition synthesis component to provide the condition for the
            boundary case.
        }{
            ACS can only tackle simple boundary cases, 
            such as comparison with constants. Since this strategy is usually used along with
            boundary identification, therefore, when complex boundaries cannot be correctly
            identified by the approach, the repair will fail as well, such as the following boundary case ({\it Lang-2}).
            As a result, to better utilize this strategy, a powerful boundary identification mechanism is needed.
        }
    \end{itemize}
\begin{lstlisting}[style=java, numbers=none]
  +  if(str.contains("#")){
  +    throw new IllegalArgumentException("Invalid locale format: " + str);
  +  }
\end{lstlisting}

\stg{Replace an identifier with a similar one}
    \begin{itemize}
        \item \tech \revise{Within our
        knowledge, this strategy has not been adopted by existing work.}{}
       Though some approaches exist to replace
        variables~\citep{long2015staged} or
        methods~\citep{long2015staged,kim2013automatic}, similarity between
        names are not considered. 
        \item \improve{Utilize name similarity when replacing identifiers.}\\
        If we do not consider the similarity between names and replace
        variables arbitrarily, incorrect patches are likely to be generated.
        \revise{}{For example, the following code is the repair
            for defect {\it Chart-7}. From the patch, we can seen the variable \code{minMiddleIndex}s
            are replaced by \code{maxMiddleIndex}s. If we do not consider the similarity between
            their names, there are several other alternatives, such as 
            \code{start}, \code{maxStartIndex} and \code{middle}, etc., which may lead to incorrect patches.
            Others cases are similar, i.e., replacing \code{RegularTimePeriod.DEFAULT\_TIME\_ZONE} with \code{zone} in
            {\it Chart-8} and replacing \code{pt} with \code{pos} in
            {\it Lang-6}. Therefore, we need to find a proper way to
            measure name similarity and incorporate that into the
            replacing templates.
          }
    \end{itemize}

  \begin{lstlisting}[style=java, numbers=none]
    if(this.maxMiddleIndex>=0){
  -   long s=getDataItem(this.minMiddleIndex).
  -                          getPeriod().getStart().getTime();
  -   long e=getDataItem(this.minMiddleIndex).
  -                          getPeriod().getEnd().getTime();
  +   long s=getDataItem(this.maxMiddleIndex).
  +                          getPeriod().getStart().getTime();
  +   long e=getDataItem(this.maxMiddleIndex).
  +                          getPeriod().getEnd().getTime();
      long maxMiddle=s+(e-s)/2;
      if(middle>maxMiddle){
        this.maxMiddleIndex=index;           
      }
    }
  \end{lstlisting}
  
\stg{Compare test executions}
    \begin{itemize}
        \item \tech \revise{
      Delta debugging~\citep{zeller2002isolating} and spectrum-based fault
      localization~\citep{jones2002visualization} are loosely related to
      this strategy as they compare failed test execution and passed
      test execution to localize faults, but their methods of comparison are very
      different from the human developer.
      Furthermore, no patch generation
      approach within our knowledge has adopted this strategy.    
    }{As far as we know, there is no patch generation
        approach has utilized this strategy. A loosely related research
        work is Statistic debugging~\citep{liu2006statisticaldebugging,chilimbi2009holmes}.}
        \item \improve{\revise{}{generate a patch from the invariants}} \\
        \revise{}{Statistical debugging is similar with ours since both of them 
        try to build invariant detection models based on the test execution
        information. However, Statistical debugging is a fault localization
        approach, thus it cannot generate patches like we do, which
        calls for more accurate models to correctly identify invariants
        related to the test failure.}
    \end{itemize}

\stg{Interpret comments}
    \begin{itemize}
        \item \tech Some approaches have adopted natural
        language processing techniques to analyze comments and other
        documents in a natural language. For example,
        ACS~\citep{xiong2016precise} analyzes the Javadoc to exclude unlikely
        variables in an ``if'' condition, and R2Fix~\citep{liu2013r2fix}
        generates patches by analyzing the bug reports in a natural
        language.
        \item \improve{Incorporate dynamic information of test failure.}
            \\
        The depth of automatic analysis still cannot
        match that \revise{of a human developer.}{in our analysis.}
        \revise{}{
            For example, the following patch is generated to fix the defect of {\it Closure-9}
            based on the comments in our analysis. 
            For current automatic techniques, it is impossible 
            to interpret this comments to the corresponding
            source code. Moreover, even though they can parse the 
            natural language, it may be confused about which character 
            should be replaced. Therefore, we need to associate the comments with
            the runtime information. By running the test cases, we can find that
            only the failed test input contains character ``/", 
            while all passed test inputs contain character ``$\setminus$", 
            which provides us the implication of replacing
            the character ``/" with ``$\setminus$". 
            As a result, 
            more robust natural language understanding is imperative.
            Besides, incorporating the dynamic information with the natural
            language understanding is needed as well.
        }
    \end{itemize}

\begin{lstlisting}[style=java, numbers=none]
    //The DOS command shell will normalize "/" to "\", 
    //so we have to wrestle it back.
  + filename = filename.replace("\\", "/");
\end{lstlisting}  
  
\stg{Imitate similar code element}
    \begin{itemize}
        \item \tech A related strategy 
        adopted by several automatic program repair approaches~\citep{le2012genprog,weimer2013leveraging,xiong2016precise} is to copy
        code pieces from other parts of the program to the potentially
        faulty location to generate patches. \revise{However, so far these approaches
          only copy the code pieces, and do not adapt the code piece for the
          new task as the human developer does.}{}
        \item \improve{\revise{}{Properly adapt the related code pieces}}\\
        \revise{}{These related approaches
        only copy the code pieces, and do not adapt the code piece for the
        new task as we do.
            For example, the following code is the patch for defect {\it Chart-2}, which is impossible
            to be generated by current automatic repair techniques as far as we know. However, the
            very similar code snippet exists nearby in the same file,
            which is listed at the below of the patch. 
            Therefore, in our analysis, we can successfully generate this patch by imitating the referred code snippet
            with imperative adaptation operations.
            However, it is impossible for current techniques to generate this correct
            patch since they simply reuse those existing code snippets without necessary adaptations for the new task.
            In this example, we need to replace some incompatible identifiers and insert additional constant comparisons,
            for which we need to accurately identify the correspondence relations among those identifiers.
        }
    \end{itemize}

\begin{lstlisting}[style=java, numbers=none]
    //patch for defect chart-2
    for(int item=0;item<itemCount;item++){
  +   if(minimum==Double.POSITIVE_INFINITY 
  +    &&maximum==Double.NEGATIVE_INFINITY
  +    &&!Double.isNaN(intervalXYData.getXValue(series,item))){
  +     minimum=intervalXYData.getXValue(series,item);
  +     maximum=minimum;
  +   }
      lvalue=intervalXYData.getStartXValue(series,item);
      uvalue=intervalXYData.getEndXValue(series,item);
      if(!Double.isNaN(lvalue)){
        minimum=Math.min(minimum,lvalue);
  +     maximum=Math.max(maximum,lvalue);
      }
      if(!Double.isNaN(uvalue)){
  +     minimum=Math.min(minimum,uvalue);
        maximum=Math.max(maximum,uvalue);
      }
    }
    //existing code snippet that is similar to the faulty code
    for(int row=0;row<rowCount;row++){
      for(int column=0;column<columnCount;column++){
        value=icd.getValue(row,column);
        double v;
        if((value!=null)
           &&!Double.isNaN(v=value.doubleValue())){
          minimum=Math.min(v,minimum);
          maximum=Math.max(v,maximum);
        }
        lvalue=icd.getStartValue(row,column);
        if (lvalue!=null
           &&!Double.isNaN(v=lvalue.doubleValue())){
          minimum=Math.min(v,minimum);
          maximum=Math.max(v,maximum);
        }
        uvalue=icd.getEndValue(row,column);
        if(uvalue!=null 
           &&!Double.isNaN(v=uvalue.doubleValue())){
          minimum=Math.min(v,minimum);
          maximum=Math.max(v,maximum);
        }
      }
    }
\end{lstlisting}


\revise{
From the above analysis we can see that, although many of the
strategies have been considered in existing approaches, still some of them
(e.g. {\it Replace an identifier with a similar one})
have not been considered by any approaches, and some of them (e.g.
{\it Interpret comments}) are not
applied in the same way or in the same depth as the human developer
does. As a result, we believe by automating these strategy we can
improve the performance of current fault localization and patch
generation approaches.}{}

\revise{}{
From the above analysis we can see that, although many of the
strategies have been considered in existing approaches, still some of them
(e.g. {\it Replace an identifier with a similar one})
have not been considered by any approaches, and some of them (e.g.
{\it Imitate similar code element}) are not
applied in the same way or in the same depth as we do. 
}

\finding{While existing techniques have already explored strategies
  similar to some of the strategies we identified, they have potential
  to be further improved based on the identified strategies.}

More importantly, many of the current approaches only utilize a single
strategy to localize or repair defects. However, as our results show,
no single strategy can be effective on a large portion of the defects.
Furthermore, most of the defects require multiple strategies to
localize and to repair. 
\revise{}{
For instance, to correctly locate the faulty code of {\it Lang-1},
we not only use the {\it Stack trace analysis} but also
{\it Excluding unlikely candidates} strategy. Furthermore,
we can notice that both of the defects explained in strategies~\ref{stg:compExecution} 
and~\ref{stg:comment} applied strategy {\it Return expected output} besides the strategy
explained for each.}
This observation calls for the studies on
combining different fault localization and patch generation approaches.


\finding{No strategy can handle all defects. Combinations of
  strategies are needed to repair a large portion of defects. }
\section{Discussion}
\label{sec:discussion}
\revise{
In this section we discuss the generalizability of our results. First,
our results are obtained by examining the defect repair process of
only one developer. Thus, our result should not be interpreted as an
understanding of the general manual repair process, as the repair
processes between different developers are likely to be  different. On
the other hand, the strategies we obtained from the process are valid,
as their effectiveness has been evaluated on the 50 defects.}{}

\revise{
Second, the case study involves 50 defects and 5 projects, and they
may not be representative for a wide range of defects in different
types of project. As a result, our results on the effectiveness of the
strategies may not be generalizable to a wider range of projects. On
the other hand, the defects are obtained from
Defects4J~\citep{just2014defects4j}. This benchmark is widely-used in
evaluating different approaches, and so far no generalizability issue
of these results is reported. Furthermore, we evenly sampled the
defects among the 5 projects. These facts give us a reasonable degree of
confidence on the generalizability of our results.}{}

\revise{}{
In this section we discuss issues related to the validity of our results.}

\revise{}{
First,  we discuss the generalizability of our results. Since the 
case study only involves 50 defects and 5 projects, they
may not be representative for a wide range of defects in different
types of projects. As a result, our results on the effectiveness of the
strategies may not be generalizable to a wider range of projects. On
the other hand, the defects are obtained from
Defects4J~\citep{just2014defects4j}. This benchmark is widely-used in
evaluating different approaches, and so far no generalizability issue
of these results is reported. Furthermore, we evenly sampled the
defects among the 5 projects and the effectiveness of those strategies 
has been evaluated on them. These facts give us a reasonable degree of
confidence on the generalizability of our results.
}

\revise{}{
Second, even though we have no prior knowledge about those defects
to be analyzed, some basic insights about those projects can be implicitly
obtained along with the analysis going on, which may cause training effect to
the subsequent analysis. As a result, when summarizing the defects
requiring the two program understanding strategies, we may
accidentally miss some defects as the program understanding happened 
unintentionally. To avoid this problem, we have carefully reviewed the
analysis record to ensure that the rest of the defects can be fixed without
program understanding. Please also not that the validity of the
main findings, including the strategies and improvements suggested to
existing techniques, are not affected by the threat.
%
%
%
}

\revise{}{
Third, as also mentioned in the introduction, our results should not be
interpreted as an upper bound of the performance of automatic program 
repair techniques since they may be superior to human developers on some
aspects as well, e.g., by utilizing its computation power. 
In other words, our results show what automatic techniques can
potentially do, but not what they cannot do.
}

\revise{}{
Fourth, our study should not be interpreted as an understanding of how
human debugs. The setting of our analysis is different from general
human debugging and a single analysis session is not enough to answer such a
question. In the related work section we have summarized some related work on
that problem.
}

\section{Conclusion and Future Work}
\label{sec:conclusion}

In this paper, \revise{we have empirically studied a manual repair process of
50 defects.  We measured the performance of the human developer,
}{
we analyzed 50 real world defects to identify to what extent these
defects can be fixed with weak test suites, based on which we}
summarized the fault localization and patch generation strategies used
\revise{by the human developer}{in our analysis}, and discussed the potential of these
strategies to be automated to improve automatic program repair.

Our findings suggest that \reviseno{}{
    most of these defects can be fixed in our analysis even 
    though without complete specifications and }there is potentially a lot of room for
current techniques to improve, and the strategies we identified could
potentially be automated and combined to improve the performance of
automatic program repair. These findings call for future work on the
automation of the strategies and the combination of the automated
strategies, leading to better automatic
program repair techniques.

\bibliographystyle{spbasic}      


\bibliography{ref}

\end{document}